\documentstyle[12pt]{article}
\def\NPB{{\it Nucl. Phys. }{\bf B}}
\def\PL{{\it Phys. Lett. }}
\def\PRL{{\it Phys. Rev. Lett. }}
\def\PRD{{\it Phys. Rev. }{\bf D}}

\def\JMP{{\it J. Math. Phys. }}

\def\IJMPA{{\it Int. J. Mod. Phys. }{\bf A}}
\def\MPL{{\it Mod. Phys. Lett. }}

\def\ie{{\it i.e.},}

\newcommand{\eq}{\begin{equation}}
\newcommand{\en}{\end{equation}}
\newcommand{\eqn}{\begin{eqnarray}}
\newcommand{\enn}{\end{eqnarray}}
\newcommand{\nn}{\nonumber }
\newcommand{\beq}{\begin{equation}}
\newcommand{\eeq}{\end{equation}}

\let\a=\alpha

\let\b=\beta
\let\d=\delta

\def\cef{c}%{c_{\rm eff}}
\def\const{\hbox{\it const.\/}}
\def\CP#1{\relax\ifmmode\IP^{#1}\else\IP$^{#1}$\fi}
\def\rd{{\rm d}}
\def\define{\buildrel{\rm def}\over=}

\def\enh{enhan\c{c}on}

\def\cE{{\cal E}}
\let\F=\Phi

\def\inv#1{{\textstyle{1\over#1}}}
\def\IP{\relax\leavevmode{\rm I\kern-.18em P}}

\def\cLef{{\cal L}}%{{\cal L}_{\rm eff}}

\def\Ione{\relax\leavevmode{\rm 1\kern-3pt l}}
\def\mef{m}%{m_{\rm eff}}
\let\q=\theta

\let\p=\pi
\def\re{r\mkern-2mu_{\rm e}}
\def\ri{r\mkern-1mu_{\rm i}}

\def\rr{r\mkern-2mu_{\rm r}}
\def\rs{r\mkern-2mu_{\rm s}}
\def\ro{r\mkern-2mu_{\rm o}}

\let\t=\tau
\def\cT{{\cal T}}
\let\To=\Rightarrow

\def\Vef{{\cal V}}%{V_{\rm eff}}
\let\vd=\partial
\let\W=\Omega

\def\IR{\relax\leavevmode{\rm I\kern-.18em R}}
\def\ZZ{\relax\leavevmode
             \ifmmode\mathchoice
             {\hbox{\sf Z\kern-.4em Z}}
             {\hbox{\sf Z\kern-.4em Z}}
             {\lower.9pt\hbox{\scriptsize\sf Z\kern-.36em Z}}
             {\lower1.2pt\hbox{\tiny\sf Z\kern-.36em Z}}
              \else{\sf Z\kern-.4em Z}\fi}
\def\RR{\relax\leavevmode
             \ifmmode\mathchoice
             {\hbox{\sf R\kern-.4em R}}
             {\hbox{\sf R\kern-.4em R}}
             {\lower.9pt\hbox{\scriptsize\sf R\kern-.36em R}}
             {\lower1.2pt\hbox{\tiny\sf R\kern-.36em R}}
              \else{\sf R\kern-.4em R}\fi}

\let\ttt=\textstyle

\catcode`@=11   % WATCH OUT !!!!! This must be reset in the end !!!!!
\def\resetby#1#2{\@addtoreset{#2}{#1}}
\def\seceq{\@addtoreset{equation}{section}% Numbers Eq.s within Sect.s
       \def\theequation{\thesection.\arabic{equation}}} % (Sect.Eq)
\catcode`@=12                   % You see ?

\def\Label#1{\label{#1}%
   \smash{\hbox to0pt{\raise1ex\hbox{\tiny[#1]}\hss}}}
\def\noLabels{\let\Label=\label}

\thicklines     

\input epsf.tex
\seceq
\noLabels % uncomment for final production
\begin{document}

\begin{titlepage}
\begin{flushright}
CITUSC/00-065\\
hep-th/0012180\\
\end{flushright}

\begin{center}

{\Large\bf\uppercase{
        On Relativistic Brane Probes\\[3mm]
        In Singular Spacetimes
           }}\\[10mm]
{\bf P. Berglund\footnote{e-mail: berglund@citusc.usc.edu} } \\[1mm]
        CIT-USC Center for Theoretical Physics\\
       Department of Physics and Astronomy\\
       University of Southern California\\
       Los Angeles, CA 90089-0484\\[5mm]
{\bf T. H\"{u}bsch\footnote{e-mail: thubsch@howard.edu}%
       $^,$\footnote{On leave from the ``Rudjer Bo\v skovi\'c'' Institute,
                 Zagreb, Croatia.} } \\[1mm]
       Department of Physics and Astronomy\\
       Howard University\\
       Washington, DC 20059\\[5mm]
{\bf D. Minic\footnote{e-mail: minic@citusc.usc.edu} } \\[1mm]
       CIT-USC Center for Theoretical Physics\\
       Department of Physics and Astronomy\\
       University of Southern California\\
       Los Angeles, CA 90089-0484\\[10mm]
\begin{flushright}\small\sl
     The brain is drained\\[-1.5mm]
     And a brown ament, and the noun I meant\\[-1.5mm]
     To use but did not, dry on the cement.\\[-1mm]
     --- John Francis Shade
\end{flushright}

{\bf ABSTRACT}\\[3mm]
\parbox{5.0in}{
We study the relativistic dynamics of brane probes in singular warped
spacetimes and establish limits for such an analysis. The behavior
of the semiclassical brane probe wave functions implies that unitarity
boundary conditions can be imposed at the singularity.}
\end{center}

\end{titlepage}

\section{Introduction}
\Label{s:IRS}
Various brane configurations in string theory
describe spacetime geometries which possess repulsive
singularities~\cite{KL}. These repulsons can be
resolved in string theory via the \enh\ mechanism~\cite{joep, alex, cliff,
clifford, cliffordII, lerda}, which describes the constituent branes that
effectively repel each other, thus forming a nonsingular matter shell of
finite radius. The original \enh\ of Ref.~\cite{joep} involved wrapping $N$
$Dp$ branes on a $K3$ manifold, the physics of which is described by large
$N$ gauge theories with eight supercharges. The naive singular repulson
geometry is replaced by a non-singular configuration of finite radius
$r_e$, where the tension of the wrapped branes vanishes.
This mechanism is related to the appearance of a negative number of  induced
$(p{-}4)$-branes upon wrapping the $p$-branes on the
$K3$ manifold~\cite{bsv}.

Recently, we have discussed an analogue of the \enh\ mechanism in a
non-supersym\-metric set-up~\cite{bhm2}, which involved space-time varying
string vacua~\cite{vafa, gh} with exponentially
large hierarchy~\cite{bhm1, cohen}. Continuing this work, in this note
we use the Dirac-Born-Infeld  action~\cite{BI, joed} to describe a
relativistic  probe analysis of a class of singular spacetime geometries
including  both repulsive and attractive naked singularities.
We discuss the familiar supersymmetric (BPS) examples, and certain
non-supersymmetric generalizations~\cite{bhm1,dudas,ZhZh,lerdaII}.
  Furthermore, we find that the behavior of the semiclassical brane probe wave
functions suggests that unitarity boundary conditions can be imposed at the
singularity. We briefly discuss the relation of this result to various
prescriptions for the resolution of singularities~\cite{singone, singtwo}.
  Our analysis is also consistent with the generic resolving properties of
$D$-brane probes~\cite{dkps, stu}.

Throughout, we consider spacetime geometries where the metric is constant
in the $p$ `longitudinal' (brane) directions and varies over the
`transverse' space, but in a `spherically' symmetric way. That is, we
assume the spacetime metric---{\it in the Einstein frame\/}---to be of the
type
\begin{equation}
       \rd s_E^2 = e^{2 A(r)} \eta_{ab} \rd x^a \rd x^b
       + e^{2B(r)} (\rd r^2 + r^2\rd\vec\W_{8-p}^2),\Label{e:metric}
\end{equation}
where $a,b=0,{\cdots},p$ and $\rd\vec\W$ is the transverse `spherical'
angle element. The explicit forms of the warp factors in various applications
are listed in section~\ref{s:Appl}.

\section{Relativistic brane probe dynamics}
\Label{s:BrPGeom}
In this section, we will derive a band of allowed energies for relativistic
$Dp$-brane probes in warped singular spacetimes, and a WKB estimate for their
wave-functions. These wave-functions will be used to argue the existence of
unitarity boundary conditions at the singularity.

The action functional that is used to determine the dynamics of a
$Dp$-brane probe in a given background is the Dirac-Born-Infeld
action~\cite{BI, joed}:
\begin{equation}
     S_p~=~-\t_p\int\rd^{p+1}\xi~e^{-\F}
         \sqrt{\det[G^s_{ab}+B_{ab}+2\p\a'F_{ab}]}
           +\mu_p\int\rd^{p+1}\xi~C_{p+1}~,
   \Label{e:BI}
\end{equation}
where $\t_p=\mu_pg_s^{-1}$ and $\mu_p=2\p(2\p\sqrt{\a'})^{-(p+1)}$
are the brane probe's tension and charge, respectively.

Aligning the brane coordinates with the spacetime ones, $\xi^0=t$
and $\xi^a=x^a$, $a=1,{\cdots},p$,
induces the metric
on the brane probe from that on the spacetime. For diagonal spacetime
metrics~(\ref{e:metric}), the metric induced on a $Dp$-brane is, in the
string frame,
\begin{equation}
     [G^s_{ab}]~=~
      \hbox{\rm diag}[\,-(e^{2A+\F/2}{-}e^{2B+\F/2}v^2)\,,\,
                       \underbrace{e^{2A+\F/2},\cdots,e^{2A+\F/2}}_p\,]~.
       \Label{e:IndBrMet}
\end{equation}
Furthermore, herein we focus on cases with $B_{ab}=0=F_{ab}$, which
simplifies the action~(\ref{e:BI}) to that of a relativistic `particle':
\begin{equation}
    S_p~=~\int\rd^{p+1}\xi~\cLef~,\qquad
    \cLef~=~-\mef\,\cef^2\sqrt{1-{v^2\over\cef^2}}-\Vef~,
    \Label{e:Leff}
\end{equation}
where
\begin{equation}
    \cef~=~e^{A-B}~,\quad\mef~=~\t_p\,e^{{p-3\over4}\F}e^{(p-1)A+2B}~,\quad
     \Vef~=~-\mu_pC_{p+1} \Label{e:cmV(r)}\quad
\end{equation}
all depend on $r$, and can be interpreted as the `effective' speed of
light, mass and potential for the brane probe, respectively.
   We first derive several results in this `general' notation, and then
apply these to brane probes of various codimensions.

\subsection{Limits on the brane probe's total energy}
Given the Lagrangian~(\ref{e:Leff}), the total energy is obtained through
the Legendre transform\footnote{The Legendre transform is of course
performed with respect to all `spherical' coordinates; the final result is
re-expressed in terms of the `total' speed, $v$. Throughout this note, we
will explicitly write formulae valid for the case of codimension-2;
higher codimension generalizations are easy to
generate.}:
\begin{equation}
    \cE~=~{\mef\cef^2\over\sqrt{1-{v^2/\cef^2}}}+\Vef~.
    \Label{e:Eff}
\end{equation}
Using the conservation of total energy, Eq.~(\ref{e:Eff}) can be solved to
express:
\begin{equation}
    v^2~=~\cef^2\bigg[1-{\mef^2\cef^4\over(\cE-\Vef)^2}\bigg]~.
\end{equation}
Requiring this to be positive, we obtain the lower bound on the total
energy (density):
\begin{equation}
    \cE~\geq~\mef\cef^2+\Vef
    ~=~\t_p\,e^{{p-3\over4}\F}e^{(p+1)A}-\mu_pC_{p+1}~. \Label{e:LoEff}
\end{equation}
  The `non-relativistic' (small $v/\cef$) expansion of Eq.~(\ref{e:Leff})
is\footnote{The angular momentum is conserved, and is
$\ell={\mef r^2v_\q\over\sqrt{1-v^2/\cef^2}}=\const$ for codimension-2.}
\begin{equation}
    \cLef~\approx~\inv2\mef v_r^2-(\mef\cef^2+\Vef-{\ell^2\over2\mef r^2})~.
    \Label{e:NRcLeff}
\end{equation}
Of course, when $v/\cef\sim1$, the Lagrangian~(\ref{e:NRcLeff}) is no
longer reliable. In addition, the family of Lagrangians~(\ref{e:Leff})
omit various other effects (perturbation of the background by the probe,
back-reaction, radiation, {\it etc.\/}). It seems plausible that these
effects are less and less negligible as $v/\cef\to1$, but there is no strict
critical value of
$v/\cef$ where the family of Lagrangians~(\ref{e:NRcLeff}) or even their
`exact' counterparts~(\ref{e:Leff}) should abruptly become incomplete.

However, one expects that the discrepancy between the `relativistic' and
the `non-relativistic' kinetic energy should be bounded by a multiple of
the latter:
\begin{equation}
    \cT-\cT_{NR}~\leq~\nu\cT_{NR}~, \quad\hbox{\ie}\quad
    \cT~\leq~(1{+}\nu)\,\inv2\mef v^2~,\qquad \nu>0~.
\end{equation}
Using Eq.~(\ref{e:Eff}), this requirement yields:
\begin{equation}
    (1{+}\nu)\,\inv2\mef v^2~\geq~\cE-\cE\Big|_{v=0}
    ~=~\mef\cef^2\bigg[{1\over\sqrt{1-{v^2/\cef^2}}}-1\bigg]~,
\end{equation}
which is a cubic inequality for $v/\cef$. This reproduces the $v^2\geq0$
inequality which led to~(\ref{e:LoEff}), and produces two more limiting
values one of which is negative and so unphysical. The third inequality
reads:
\begin{equation}
    {v\over\cef}~\leq~
    {\ttt\sqrt{\sqrt{(\nu+1)(\nu+9)}+\nu-3\over2(\nu+1)}}
    ~~~\left\{\begin{array}{ll}
        =0&\mbox{if~~}\nu=0~,\\%[1mm]
        \approx0.786&\mbox{if~~}\nu=1~,\\%[1mm]
        =1&\mbox{if~~}\nu=\infty~.
              \end{array}\right. \Label{e:UpVcty}
\end{equation}
Clearly, the `ultra-relativistic' case implies an infinite discrepancy
between the `relativistic' and `non-relativistic' kinetic energies.
   We will hereafter use the $\nu=1$ limit.
   When inserted in Eq.~(\ref{e:Eff}), this produces an {\it upper\/} limit
on the total energy (density):
\begin{eqnarray}
    \cE&\leq&\b\,\mef\cef^2+\Vef
    ~=~\b\,\t_p\,e^{{p-3\over4}\F}e^{(p+1)A}-\mu_pC_{p+1}~,
     \Label{e:HiEff}\\[1mm]
    \b&=&{\ttt\sqrt{2(\nu{+}1)\over\nu{+}5-\sqrt{(\nu{+}1)(\nu{+}9)}}}
          \approx1.618\quad\hbox{for}\quad\nu=1~. \Label{e:beta}
\end{eqnarray}

Thus, the total energy (density) is bracketed as
\begin{equation}
    \mef\cef^2+\Vef~\leq~\cE~\leq~\b\,\mef\cef^2+\Vef~. \Label{e:HiLoEff}
\end{equation}
This is the main result of the classical brane probe analysis. This
extended inequality will turn out to `screen' the singularities from the
brane probe. Explicit examples will be given in section~\ref{s:Appl}.

\subsection{Semiclassical analysis}
\Label{s:WKB}
The radial momentum may be rewritten as
\begin{equation}
    \wp_r~\define~{\vd\cLef\over\vd v_r}
     ~=~{\mef v_r\over\sqrt{1-v^2/\cef^2}}
     ~=~{\chi\mef v_r\over\sqrt{1-v_r^2/\cef^2}}~, \Label{e:rMom}
\end{equation}
where (in codimension-2)
\begin{equation}
    \chi~\define~\bigg[1-{1\over1+{\mef^2\cef^2r^2/\ell^2}}\bigg]^{-1/2}
\end{equation}
accounts for the (constant) angular momentum. Similarly, the total energy
function~(\ref{e:Eff}) becomes
\begin{equation}
    \cE~=~{\chi\mef\cef^2\over\sqrt{1-{v_r^2/\cef^2}}}+\Vef~.
    \Label{e:cEff}
\end{equation}
Eliminating $v_r$ from Eqs.~(\ref{e:rMom})--(\ref{e:cEff}), we obtain
\begin{equation}
    -\wp_r^2+(\cE-\Vef)^2-\chi\mef\cef^2~=~0~.
\end{equation}
Upon substituting $\wp_r\to-i\hbar\vd_r$, this produces the Klein-Gordon
equation for the brane probe's (stationary) wave-function:
\begin{equation}
    \vd_r^2\psi(r)
     ~+~(\hbar\cef)^{-2}\Big[(\cE-\Vef)^2-\chi^2\mef^2\cef^4\Big]
      \psi(r)~=~0~.
\end{equation}
   For future reference, the `relativistic' WKB wave-functions for the brane
probe is
\begin{equation}
    \psi_\pm(r)~=~{1\over\sqrt{k(r)}}e^{\pm i\int\rd r\,k(r)}~,\qquad
     k(r)~\define~\inv{\hbar\cef}\sqrt{(\cE-\Vef)^2-\chi^2\mef^2\cef^4}~,
    \Label{e:relWKB}
\end{equation}
and for small $\cef$:
\begin{eqnarray}
    \cef^2\ll{(\cE-\Vef)\over\chi\mef}
    &\To&k(r)\sim{(\cE-\Vef)\over\hbar\cef}\nn\\[1mm]
    &\To&\psi_\pm(r)\sim
     \sqrt{\hbar\cef\over(\cE-\Vef)}\,e^{\pm i\varphi(r)}~.
     \Label{e:ceff0}
\end{eqnarray}

In the case of interest, $\cef^2$ and the wave-function~(\ref{e:ceff0})
both vanish at the singularity, thus preserving probability.

\section{Applications}
\Label{s:Appl}
The above general results apply straightforwardly to several examples of
recent interest, which we discuss in this section. Let us start by analyzing
the BPS solution wrapped on a
$K3$~\cite{joep,alex,cliff,clifford}~\footnote{Our notation is as follows:
$x_\|$ shall denote coordinates parallel to the unwrapped part of the brane,
and $x_\perp$ those that are transverse to the whole brane probe.}:
\begin{eqnarray}
    \rd s^2_s&=&Z_{p-4}^{-1/2}Z_p^{-1/2}\|\rd x_\|\|^2 +
    Z_{p-4}^{1/2}Z_p^{-1/2}\rd s_{K3}^2 +
Z_{p-4}^{1/2}Z_p^{1/2}\|\rd x_\perp\|^2~,\nn\\
    e^\F&=&g_s\,Z_{p-4}^{7-p\over4}Z_p^{3-p\over4}~,\nn\\
    C_{p-3}&=&(Z_{p-4}^{-1}-1)g_s^{-1}\rd^{p-3}x_\|~,\quad%\nn\\
   C_{p+1}~=~(Z_p^{-1}-1)g_s^{-1}\rd^{p-3}x_\|\wedge\rd^4x_{K3}~.
\Label{e:pcodim>2}
\end{eqnarray}
The harmonic functions $Z_p$ and $Z_{p-4}$ are
given as
\begin{eqnarray}
    Z_p~&\define&~1+a_pr^{p-7}~, \quad\quad\quad\hbox{where}\quad
    a_p~\define~N\,g_s\,(2\sqrt{\p})^{1-p}(\sqrt{\a'})^{3-p}
                 \Gamma\Big({\ttt{3-p\over2}}\Big)\nn\\
    Z_{p-4}~&\define&~1-{V_*\over V_{K3}}a_pr^{p-7}~,
    \quad\hbox{and}\quad V_*~\define~(2\p\a'^{1/2})^4~.
                 \Label{e:Zp}
\end{eqnarray}
Note that $Z_p$ ($Z_{p-4}$) is a monotonically
decreasing (increasing) function:
$Z_p(0)=+\infty$ and $Z_p(\infty)=1$, while $Z_{p-4}(\rr)=0$ and
$Z_{p-4}(\infty)=1$.
  In fact, $\rr$ is the location of a repulsive singularity~\cite{joep}.

Let us now study the brane probe dynamics following section~\ref{s:BrPGeom}.
  From Eqs.~(\ref{e:cmV(r)}), the warp factors $e^{2A}, e^{2B}$ being the
metric components along $\|\rd x_\|\|^2$ and $\|\rd x_\perp\|^2$,
and Eqs.~(\ref{e:pcodim>2}) we have
\begin{eqnarray}
    \cef~&=&~Z_{p-4}^{-1/2} Z_p^{-1/2}~, \Label{e:c}\\
    \mef~&=&\t_pZ_{p-4}^{+1}(V_{K3}-V_*Z_pZ_{p-4}^{-1})~,
     \Label{e:tension}\\
    \Vef~&=&~\t_p\big[Z_p^{-1}(V_{K3}-V_*Z_pZ_{p-4}^{-1})
    -(V_{K3}-V_*)\big]~. \Label{e:BPSpcmV}
\end{eqnarray}
This gives us the following extended inequality for the total energy
(density) of the probe
\begin{equation}
0~\leq~\cE-\t_p(V_{K3}-V_*)
  ~\leq~\t_p(\b-1) Z_p^{-1}(V_{K3}-V_*Z_pZ_{p-4}^{-1})~.
\end{equation}
The second inequality gives an $r$-dependent bound. As the probe approaches
the singularity from afar, its total energy meets the bound at $\ri$; see
the upper left corner of Fig.~\ref{f:cases}. There, the effects
omitted in the action~(\ref{e:BI}) start taking over, rendering the analysis
incomplete.
\begin{figure}[h]
        \framebox{\epsfxsize=160mm%
        \hfill~%Here be picture.
  \setlength{\unitlength}{1mm}
  \begin{picture}(160,190)(0,0)
        \put(0,0){\epsfbox{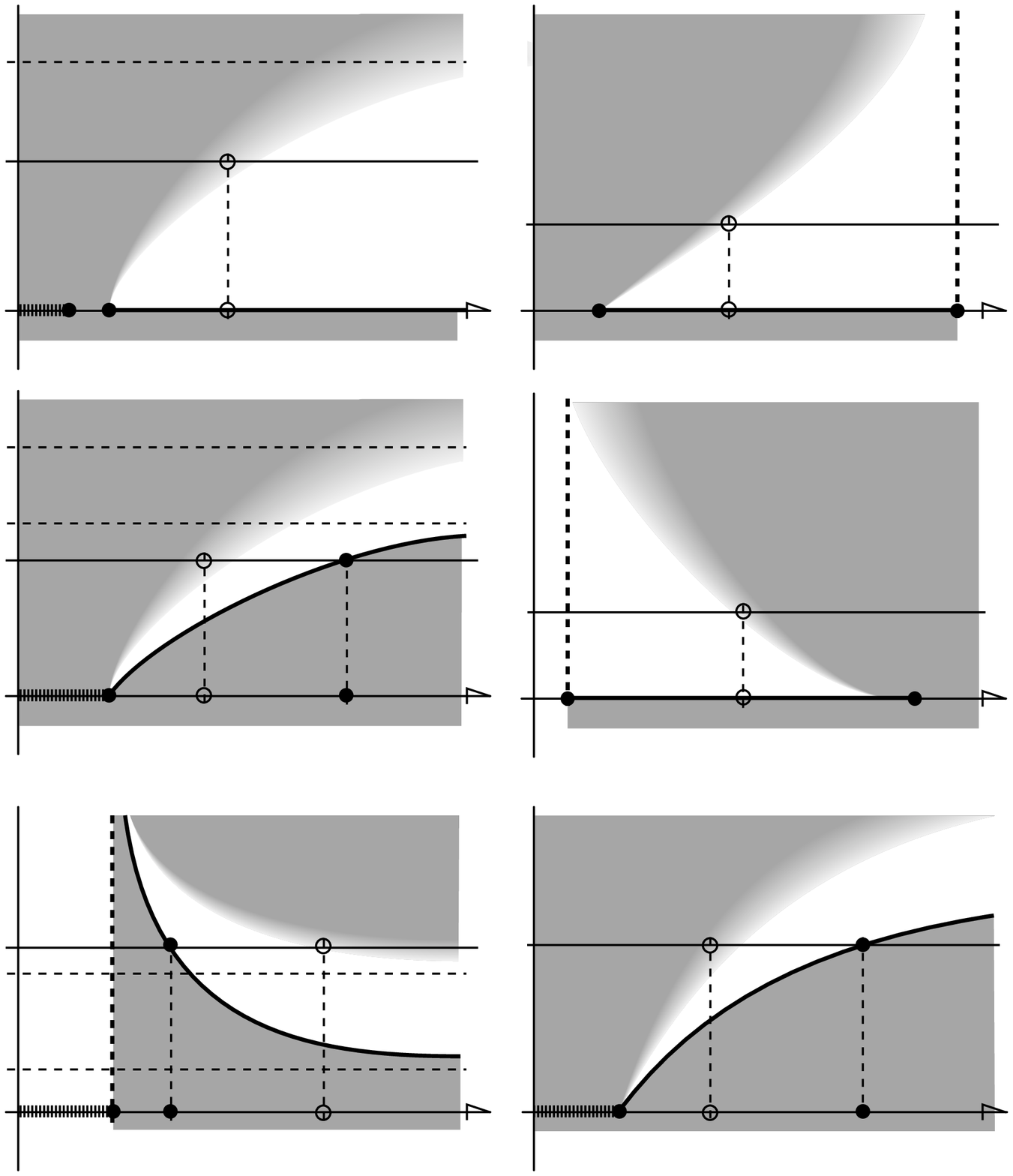}}
        \put(5,170){BPS, $p<7$}
        \put(5,165){wrapped on a $K3$}
        \put(60,162){$\cE$}
        \put(9,132){$r_{\rm r}$}
        \put(16,132){$\re$}
        \put(35,132){$\ri$}
        \put(75,132){$r$}
        \put(5,118){non-BPS, $p<7$; the `$+$' solution}
        \put(10,99){$\cE$}
        \put(16,71){$\rs$}
        \put(31,71){$\ri$}
        \put(53,71){$\ro$}
        \put(75,71){$r$}
        \put(35,53){non-BPS, $p<7$;}
        \put(35,48){the `$-$' solution}
        \put(10,37){$\cE$}
        \put(16,5){$\rs$}
        \put(26,5){$\ro$}
        \put(50,5){$\ri$}
        \put(75,5){$r$}
        \put(90,175){BPS, $p=7$;}
        \put(90,170){the `$+$' solution}
        \put(90,165){wrapped on a $K3$}
        \put(140,152){$\cE$}
        \put(95,132){$\re$}
        \put(114,132){$\ri$}
        \put(150,132){$r_\F$}
        \put(156,132){$r$}
        \put(95,115){BPS, $p=7$;}
        \put(95,110){the `$-$' solution}
        \put(95,105){wrapped on a $K3$}
        \put(140,90){$\cE$}
        \put(90,71){$r_\F$}
        \put(116,71){$\ri$}
        \put(144,71){$\re$}
        \put(156,71){$r$}
        \put(90,50){non-BPS, $p=7$}
        \put(90,37){$\cE$}
        \put(97,4){$\rs^+$}
        \put(111,4){$\ri$}
        \put(135,4){$\ro$}
        \put(156,4){$r$}
  \end{picture}
        \hfill~}
        \caption{The energy diagram for various cases of probing spacetime
singularities with branes. $\mef\cef^2$ becomes imaginary in the heavily
dashed values of $r$. At $r_\F$, $e^\F\to\infty$; $\rs,\rr$ and $\re$ are the
naked singularity, repulson and \enh\ radius, respectively.}
        \Label{f:cases}
\end{figure}

The qualitative behavior of the brane probe for codimension-2 solutions and
one class of the non-BPS solutions in codimension larger than 2 turns out to
be quite similar to the case just examined.
  In particular, in all BPS solutions, the lower energy bound in the
inequality~(\ref{e:HiLoEff}) is constant and can be set to zero. In
contrast, in non-BPS solutions, the lower energy bound depends on $r$ and
there is a corresponding limit, $\ro$, on the space in which the brane probe
can move. The upper energy bound~(\ref{e:HiLoEff}) provides a `soft' limit,
$\ri$, where the effects omitted in the action~(\ref{e:BI}) are no longer
negligible. In all but the case we examine below, this effectively `screens'
a metric singularity.

Finally we consider a non-BPS solution along the lines of
Ref.~\cite{ZhZh,lerdaII}:
\begin{eqnarray}
    \rd s^2_E&=&D_p^{{\d\over \sqrt{p+1}}}\|\rd x_\|\|^2 {+}
                D_p^{{-\sqrt{p+1}\over 7-p}\d}
     \Big[1{-}\Big({\rs\over r}\Big)^{2(7-p)}\Big]^{2\over 7-p}
      \|\rd x_\perp\|^2~,\nn\\
    e^\F,C_{p+1}&=&\const, \quad\hbox{where}\quad
    D_p~=~\bigg({1{-}({\rs\over r})^{7-p}\over
                  1{+}({\rs\over r})^{7-p}}\bigg),
    \quad\hbox{and}\quad \d~=~\sqrt{7-p\over2}~.\Label{e:p}
\end{eqnarray}
  In Fig.~\ref{f:cases}, this is labeled as the `$-$' solution. $D_p(r)$ is a
monotonically increasing function from $D_{p}(\rs)=0$ to $D_{p}(\infty)=1$.
At $r=\rs$ there is a singularity~\cite{lerdaII}.

The extended inequality on the total energy~(\ref{e:HiLoEff}) now reads:
\begin{equation}
  \t_p D_p^{\pm\d\sqrt{p+1}\over 2}\leq \cE-\Vef
   \leq \t_p\b D_p^{\pm\d\sqrt{p+1}\over 2}~.
\end{equation}
This limits the space available to the brane probe on both ends. The lower
energy limit gives a classical turning point, $\ro$, while the upper energy
limit produces the `soft' limit, $\ri$, beyond which the relativistic
effects omitted in the action~(\ref{e:BI}) can no longer be neglected.

In contrast, this brane probe analysis now is reliable near $\ro$ and $\rs$.
  From Eqs.~(\ref{e:p}), the wave function here behaves as described in
Eq.~(\ref{e:ceff0}). Across $\ro$, the standard WKB matching conditions
allow us to extend the wave function into the region $\rs\leq r\leq\ro$.
There,
\begin{eqnarray}
|\psi_\pm(r)|\sim
     \sqrt{\cef}\,e^{\mp\tilde\varphi_0\cef^\epsilon}\to0~,
  \quad\hbox{when}\quad~r\to\rs~,
\end{eqnarray}
where $\tilde\varphi_0=\const$, and $\epsilon>0$.
  Since the wave function $\psi_\pm$ vanishes at $\rs$, we can---{\it
and do\/}---set it to zero for $0\leq r\leq\rs$. As this is where
probability may leak, the vanishing of $\psi_\pm(r)$ in this region ensures
the conservation of probability.
  Our result is in agreement with the analyses in Refs.~\cite{cohen,singone}.

\section{Conclusions}
\Label{s:ABC+D}
In this note we described a relativistic probe analysis of a variety of
singular spacetime geometries governed by the general structure of the
Dirac-Born-Infeld action. In particular, we discussed the BPS solution
wrapped on a $K3$ manifold as well as certain non-BPS generalizations.
  We argued that the behavior of the semiclassical brane probe wave functions
implies that unitarity boundary conditions can be imposed at the
singularity. This is consistent with some existing prescriptions for the
resolution of singularities~\cite{singone, singtwo}. Although our analysis
is incomplete near the singularities, we note
formally that the wave functions vanish at the singularity. In addition, a
more realistic description of the brane probes should take into account
their energy loss to the singularity and radiation. We observe that the
candidate energy function
$\cE'=\cLef+\Vef$ has the right qualitative properties: (1)~$\cE'$ agrees
with $\cE$ in the nonrelativistic regime, and (2)~$\cE'$ vanishes at the
singularity. The latter property implies that $\ri$ becomes a hard limit,
\ie\ the singularity is now truly screened. Moreover, the WKB wave functions
become exponentially suppressed and vanish at the singularity.

Finally, according to the generic properties of $D$-brane probes~\cite{dkps,
stu}, the resolution of the probe is $\sim\sqrt{v/\cef}\,l_s$. From our
analysis, it then follows that $v/\cef\to1$ as the brane probe approaches the
singularity. Thus, the resolution of the probe at the singularity is always
of the order of $l_s$.

{\bf Acknowledgments:}
We thank I.~Bars, A.~Brandhuber and N.~Itzaki for useful discussions.
      The work of P.~B.\ and D.~M.\ was supported in part by  the US
Department of Energy under grant number DE-FG03-84ER40168.
      T.~H.\ wishes to thank the US Department of Energy for their
generous support under grant number DE-FG02-94ER-40854,
and the Caltech-USC Center for Theoretical Physics where this work was
completed.

\end{document}